\documentclass[printer]{aa}
\usepackage[varg]{txfonts}
\usepackage{color}
\usepackage{graphicx}
\usepackage{natbib}
\usepackage[breaklinks=true]{hyperref} 
\usepackage{multirow}
\bibpunct{(}{)}{;}{a}{}{,} 
\usepackage[switch]{lineno}

\begin{document}

\title{Decayless low-amplitude kink oscillations: a common phenomenon in the solar corona?}
\titlerunning{Decayless kink oscillations: a common phenomenon in the corona?}
\author{S. A. Anfinogentov\inst{1,2}  \and V. M. Nakariakov\inst{1,3,4} \and G. Nistic\`o\inst{1}}
\institute{Centre for Fusion, Space and Astrophysics, Department of Physics, University of Warwick, Coventry CV4 7AL, UK
\label{1}
\and Institute of Solar-Terrestrial Physics, Lermontov st., 126a, Irkutsk 664033, Russia\label{2} \email{anfinogentov@iszf.irk.ru}
\and 
{School of Space Research, Kyung Hee University,
Yongin, 446-701, Gyeonggi, Korea\label{3}}
\and
Central Astronomical Observatory at
Pulkovo of the Russian Academy of Sciences,
St Petersburg 196140, Russia\label{4}}

\date{Received \today /Accepted dd mm yyyy}

\abstract 
{We investigate the decayless regime of coronal kink  oscillations  recently discovered in the Solar Dynamics Observatory (SDO)/AIA data.
In contrast to  decaying kink oscillations that are excited by impulsive dynamical processes, this type of transverse oscillations is not connected to any external impulsive impact, such as a flare or CME, and does  not show any significant decay. Moreover the amplitude of  these decayless oscillations is typically lower than that of decaying oscillations.}
{The aim of this research is to estimate the prevalence of this phenomenon and its characteristic signatures. }
{We analysed 21 active regions (NOAA 11637--11657) observed in January 2013 in the 171~\AA~ channel of SDO/AIA. For each active region we inspected six hours of observations, constructing time-distance plots for the slits positioned across pronounced bright loops. The oscillatory patterns in time-distance plots were visually identified and the oscillation periods and amplitudes were measured. We also estimated the length of each oscillating loop.} 
{Low-amplitude decayless kink oscillations are found to be present in the majority of the analysed active regions. The oscillation periods lie in the range from 1.5 to 10~minutes.  In   two active regions with insufficient observation conditions we did not identify any oscillation
patterns. The oscillation periods are found to increase with the length of the oscillating loop.}
{The considered type of coronal oscillations is a common phenomenon in the corona. The established dependence of the oscillation period on the loop length is consistent with their interpretation in terms of standing kink waves.}


\maketitle
\titlerunning{Decay-less kink oscillations}
\authorrunning{Anfinogentov et al.}

\section{Introduction}

Kink oscillations of coronal loops are possibly the most studied and debated class of coronal oscillations.  Kink oscillations are detected as transverse, in the plane-of-sky displacements of bright or dark coronal non-uniformities \citep[e.g.][]{1999Sci...285..862N, 2014ApJ...790L...2T, 2005A&A...430L..65V}, the variation of the brightness itself as a result of the modulation of the column depth of the oscillating plasma non-uniformity by the oscillation \citep[e.g.][]{2003A&A...397..765C, 2009ApJ...698..397V}, or the periodic Doppler shift of coronal emission lines \citep[e.g.][]{1983A&A...120..185K, 2007Sci...317.1192T}. Also, kink waves can produce modulation of gyrosynchrotron emission \citep[e.g.][]{khodachenko_et_al_aa_2011, kupriyanova_et_al_solphys_2013}.
Kink waves are observed as standing \citep[e.g.][]{1999ApJ...520..880A,1999Sci...285..862N,2011ApJ...736..102A} and propagating \citep[e.g.][]{2001MNRAS.326..428W, 2007Sci...317.1192T,2007Sci...318.1580C} disturbances, with the periods ranging from several seconds to several minutes. 

Kink oscillations are interpreted as fast magnetoacoustic waves guided by field-aligned non-uniformities of the fast magnetoacoustic speed \citep[see][for recent comprehensive reviews]{2012RSPTA.370.3193D, 2014SoPh..289.3233L}. In the low-$\beta$ plasma of coronal active regions, the fast speed non-uniformity corresponds to the Alfv\'en speed non-uniformity, which can be created by a field-aligned non-uniformity of the plasma density, e.g. a coronal loop.
Because of the waveguiding effect, the wave phase speed has a value between the Alfv\'en speeds inside and outside the loop. In the long-wavelength limit, the phase speed of the kink wave is the so-called kink speed \citep{zaitsev_stepanov_azh_1982, 1983SoPh...88..179E}, and the perturbation becomes weakly compressive \citep{2012ApJ...753..111G}. 

Kink oscillations are usually observed to be the lowest spatial harmonics along the field, i.e. the global (or fundamental) modes of coronal loops, with the nodes of the displacement at the loop's footpoints and the maximum at the loop apex. In some cases, second and third spatial harmonics have been detected \citep{2004SoPh..223...77V, 2007ApJ...664.1210D, 2009A&A...508.1485V}. 
The period of the global kink mode is then determined by double the length of the loop, divided by the kink speed. 

The interest in kink oscillations is mainly connected with the possible solution of the coronal heating problem \citep[e.g.][and references therein]{2013ApJ...768..191G}, and also with coronal plasma diagnostics, i.e. magnetohydrodynamic (MHD) coronal seismology \citep[e.g.][]{2008PhyU...51.1123Z, stepanov_et_al_physusp_2012}. In particular, kink oscillations are used for estimating coronal magnetic field \citep[e.g.][]{2001A&A...372L..53N}, density stratification \citep[e.g.][]{2005ApJ...624L..57A, 2008A&A...491L...9V}, the variation of the magnetic field along the loop \citep[e.g.][]{2008ApJ...686..694R, 2008A&A...486.1015V}, and information about fine structuring \citep[e.g.][]{2008A&A...491L...9V, 2014ApJ...787L..22A}. 

Standing kink oscillations of coronal loops are observed in two regimes. In the large-amplitude, rapidly-decaying regime the displacement amplitude reaches several minor radii of the oscillating loop, and the decay time equals two to four periods of the oscillation \citep[see e.g.][for a dedicated review]{2009SSRv..149..199R}. It was recently shown that in the vast majority of cases, kink oscillations are excited by a coronal eruption that mechanically displaces loops in the horizontal direction from the equilibrium \citep{ZiNa}. The decay of the oscillations is usually associated with the phenomenon of resonant absorption that is linear coupling of the collective kink oscillation with torsional Alfv\'enic oscillations at the surface of the constant Alfv\'en speed that coincides with the phase speed of the kink wave \citep[e.g.][]{2002ApJ...577..475R, 2002A&A...394L..39G}.
The low-amplitude undamped regime of kink oscillations was discovered very recently \citep{2012ApJ...751L..27W, 2013A&A...552A..57N}. The oscillations do not damp in time and are seen for a number of cycles. Sometimes the amplitude even gradually grows \citep{2012ApJ...751L..27W}. Different loops oscillate with different periods \citep{2013A&A...560A.107A}. Moreover, the same loop was seen to oscillate in two regimes, decaying and undamped, with the same period \citep{2013A&A...552A..57N}.
All segments of the loops are seen to oscillate in phase, indicating that the oscillation is standing. The displacement amplitude is about the minor radius of the loop, which is typically lower than the amplitude of  decaying oscillations. Off-limb observations show that the oscillations are polarised in the horizontal direction. Theory of undamped kink oscillations has not been developed yet, and their relationship with the decaying oscillations is unclear. However, both, damped and undamped regimes are clearly associated with the standing kink (or $m=1$) mode of oscillating loop.

The aim of this paper is to assess the persistency of undamped kink oscillations and gain some statistical information about their properties. In Sec.~\ref{obs} we describe the observational data used in our study. In Sec.~\ref{ana} we present the analytical techniques and results obtained. In Sec~\ref{con} we summarise our findings.

\section{Observational data}
\label{obs}

\begin{figure*}
\centering
\resizebox{\hsize}{!}{\includegraphics{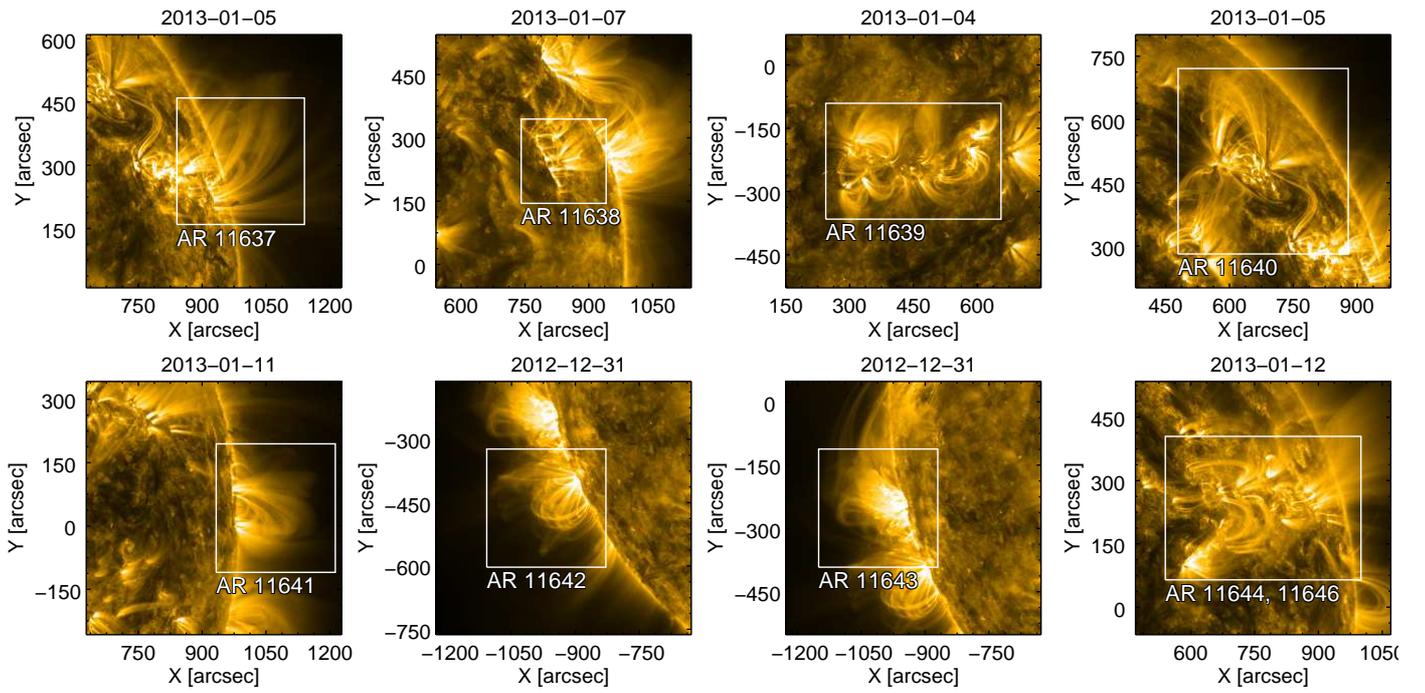}}
\caption{Examples of active regions from the observation set. The analysed active regions are enclosed by white boxes, with the specific classification numbers shown. NOAA 11644 and 11646 (right bottom panel) are located in the vicinity of each other and connected  by several coronal loops, hence,  we treated them as a single active region.}
\label{fig_1}
\end{figure*}

We analysed observations of coronal loops in a set of active regions that passed through the solar disk during about one month, between December 2012 -- January 2013. To eliminate the selection effect we analysed 21 active regions (NOAA 11637--11657) one by one. Images of some active regions, obtained in extreme ultraviolet (EUV) are presented in Fig.~\ref{fig_1}.
In this list there are very small and undeveloped active regions, e.g. NOAA 11638, as well as complex and large active regions like NOAA 11640.

\begin{figure*}
\centering
\resizebox{\hsize}{!}{\includegraphics{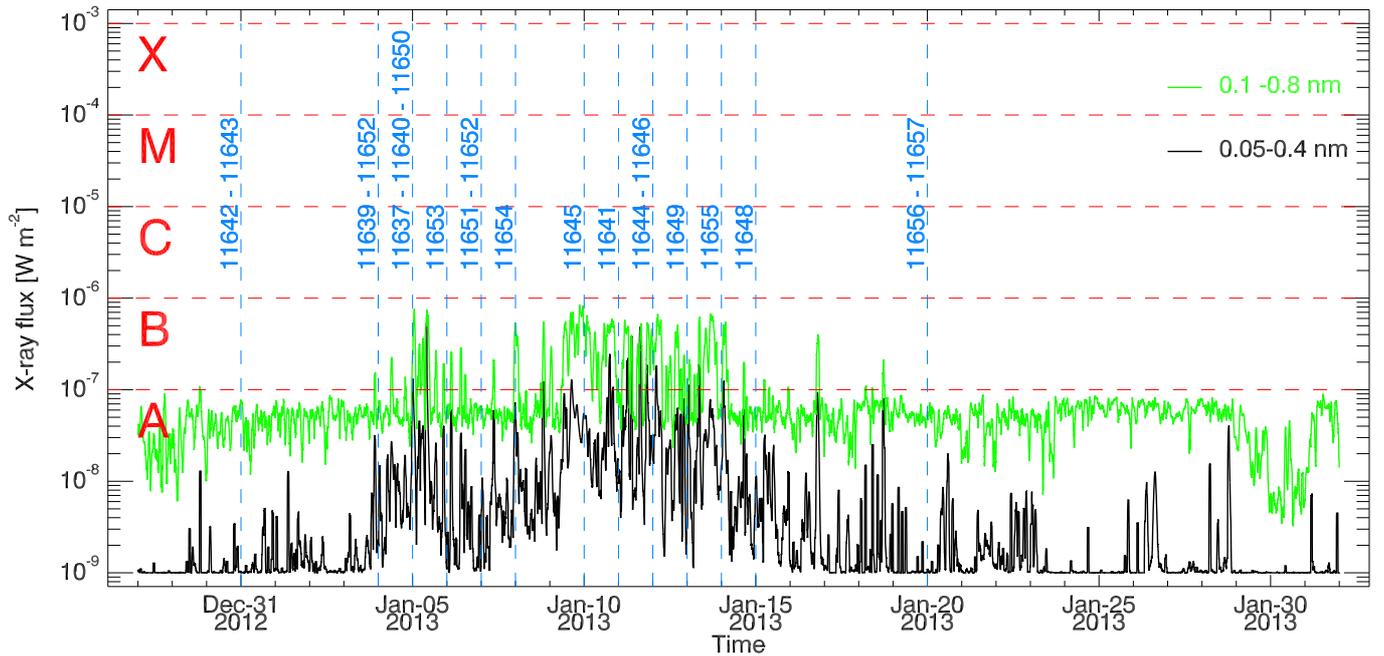}}
\caption{Total  X-ray flux measured by the GOES 15 satellite from  28 December 2012 till  2 February 2013, in the range of 0.05--0.4 nm (black) and 0.1-0.8 nm (green) smoothed over a period of one hour. The flux is measured in units of $\mathrm{W m^{-2}}$. The horizontal dashed red lines indicate the flare classes limits. The vertical dashed blue lines represent times of the observations listed in Table~\ref{table}, labelled with the identification numbers of the analysed active region.}
          \label{fig_2}
 \end{figure*}

For all of the selected active regions we downloaded  six hours of images obtained  with the Solar Dynamics Observatory (SDO)/AIA at 171~\AA, starting from 00:00~UT of the specific day of the observation, which is listed in the second column of Table~\ref{table}. 
The observation times were selected to find active regions on the solar limb or close to it, in order to analyse loops well contrasted by the darker background. Some of active regions, e.g. NOAA 11639, were analysed during their presence on the disk, as they were not visible at the solar limb as  they are overlapped by other active regions. We also excluded the time intervals that included impulsive events, such as flares and coronal mass ejections. Figure~\ref{fig_2} shows the soft X-ray flux of the Sun for the analysed period of time. It is evident that there were no solar flares stronger than the C-class, during the whole considered period of time.

The data in FITS format were retrieved from the JSOC data centre \url{http://jsoc.stanford.edu/ajax/lookdata.html} with the  highest possible spatial (0.6~arcsec) and time resolution (12~s). 
                 
\section{Analysis and results}
\label{ana}

Decayless kink oscillations of coronal loops are characterised by a very low displacement amplitudes (lower than 1~Mm, on average about 0.2~Mm, which is comparable to the SDO/AIA pixel size or less). Therefore, they are hard to see in animations. However they can be readily identified in time-distance plots made across the oscillating loop as characteristic wavy patterns. Thus, we examined time-distance plots in the chosen time intervals for all of the loop-like structures in each active region of our list. We visually looked for the presence of transverse oscillations, and when an oscillation was found, we estimated its periods and amplitudes, and the length of the oscillating loop.
Further we describe the data preparation and analysis procedure in detail.
        
\begin{figure}
    \centering
      \resizebox{\hsize}{!}{\includegraphics{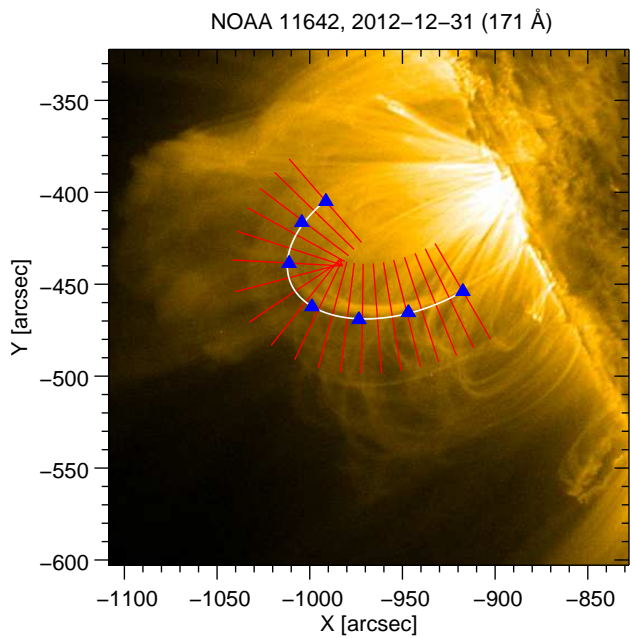}}
      \caption{EUV image of the NOAA 11642 active region. The blue triangles show the manually selected points of the analysed loop, which is fitted with an ellipse (white curve). Transverse slits perpendicular to the apparent loop path are shown with  red lines. The slits were used for making time-distance plots. For the purpose of clarity, we show only 20 slits out of 100 analysed.}
   \label{fig_3}
 \end{figure}

As the first step, we visually identified distinct coronal loops in the EUV images.
Then, we manually specified 5--10 points along the selected coronal loop.
Their coordinates were used to fit the projected loop shape with an elliptic curve in the image plane. 
We stacked 100 equidistant slits that were locally perpendicular to the loop  to depict oscillations at different positions (Fig. \ref{fig_3}). Each slit was 5-pixels wide and 100 pixels long. To increase the signal-to-noise ratio, we calculated the average intensity over the slit width. For faint loops we used wider slits of nine pixels.
         
The time-distance maps were then visually inspected,  searching for oscillatory patterns and selecting the events in which at least three cycles of the oscillation were clearly seen by eye and the displacement amplitude did not change significantly from period to period.
For the detailed analysis, we selected one slit out of 100 where the oscillation patterns are most pronounced.
We applied this procedure to all analysed loops. 
                
In EUV images, active region loops are usually seen to overlap with many other loops. Thus it is difficult to determine the transverse shape of a single loop, e.g. by a best fit with a Gaussian profile, and determining its evolution in a time-distance plots. Because of that it is more convenient to track the oscillations by the loop edges, e.g. by calculating partial derivatives of the intensity in the direction across the loop. Taking that the kink oscillation does not change the loop's minor radius, we determine the loop location each instant of time by fitting the spatial derivatives of  the transverse profile of the EUV intensity across the loop with a Gaussian function. 

\begin{figure*}
           \centering
           \resizebox{\hsize}{!}{\includegraphics{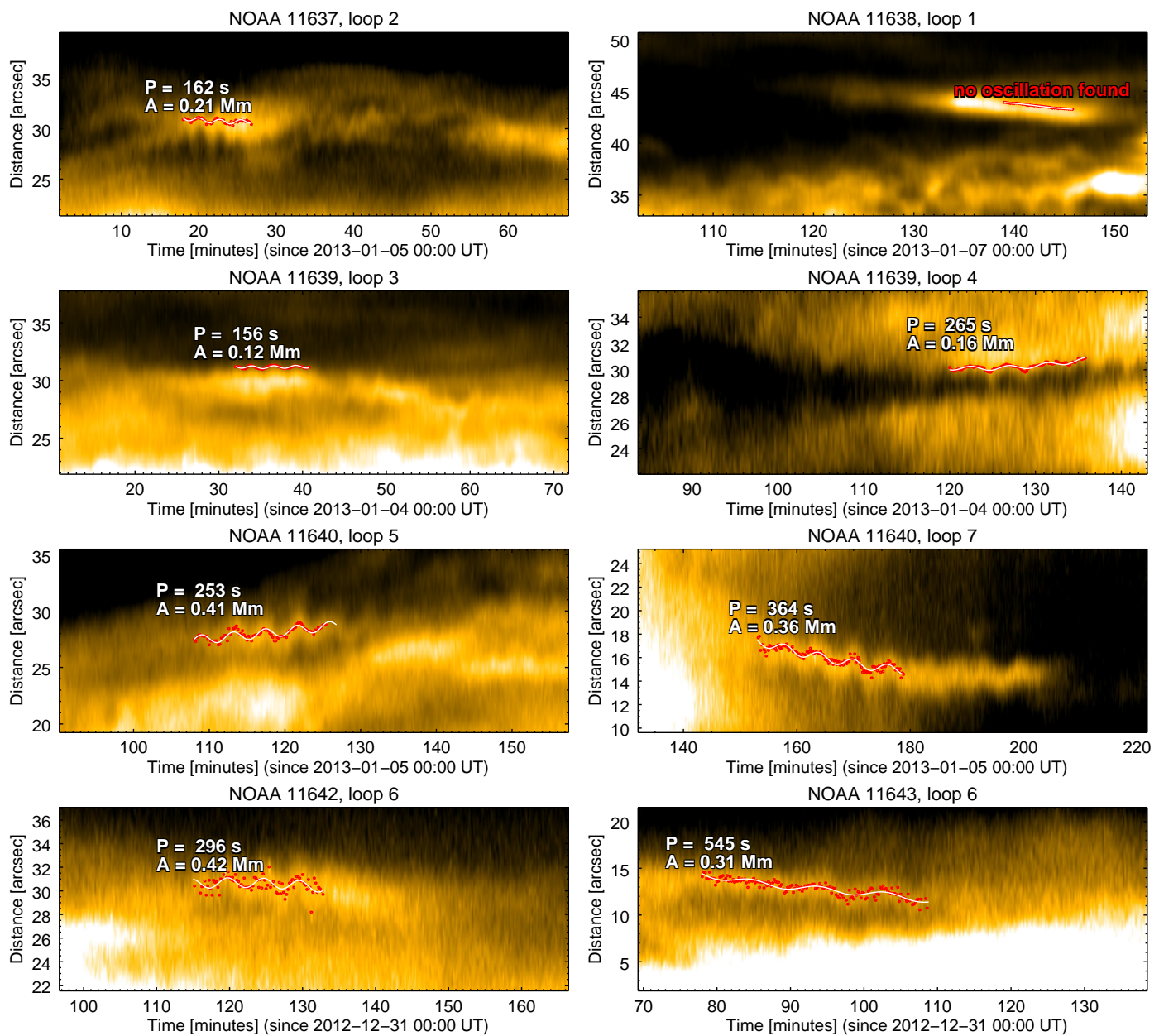}}
           \caption{Time-distance maps of the oscillating loops found in the analysed active regions. The most noticeable oscillations are fitted with a sine function to define their period and amplitude. Red dots indicate the positions of the loop centres estimated by the Gaussian fitting. The white curves show the best-fitting sinusoidal functions. In the panels we indicate the periods and amplitudes of the detected oscillations.}
 \label{fig_4}
\end{figure*}

Since the estimated amplitude of the decayless oscillation was found to be about 0.2 Mm \citep{2013A&A...560A.107A}, which is less than the SDO/AIA pixel size, the loop location must be measured with  accuracy better than one pixel. This is possible because even the  displacement of less than one pixel causes changes in the  intensity of the individual pixels across the loop. The Gaussian fitting approach allows us to recover the loop position with sub-pixel accuracy from the intensity distribution over the image pixels. The oscillation amplitude and period were measured with a best fit of the loop locations at each moment of time with a sum of a sine function and a linear trend
\begin{equation}\label{eq:sine_fit}
        x(t)=\xi\sin\left(\frac{2\pi t}{P}+\phi_0 \right) +  a_0 + a_1 t .
\end{equation}
Here $\xi$ is the oscillation amplitude, $P$ is the period, $\phi_0$ is the initial phase, $a_0$ and $a_1$ are constant values.
We do not include any decay term in Eq. (\ref{eq:sine_fit}) because   various amplitude evolutions are possible.
The oscillations can decay, grow, remain constant, or show more complex behaviour;
see e.g. the upper panel of the left column in Fig. \ref{fig_4}, where the oscillation increases from 12 to 20 min. and then decays. 

Examples of the time-distance maps with the positions of the loop edges determined by this method are shown in Fig.~\ref{fig_4}. The results of oscillation fitting with the function from Eq. (\ref{eq:sine_fit}) are shown with the  overplotted white lines. 

Low-amplitude kink oscillations were detected in all analysed active regions except NOAA 11638 and 11647. The NOAA 11638 was a small active region with short coronal loops observed when it was on the solar disk. Under these observational conditions, it is very hard to identify any oscillation because of dynamical background and shortness of the expected periods of standing kink oscillations. Also, the lack of oscillations of the bright loop, shown in the top-right panel of Fig.~\ref{fig_4}, can, e.g. be caused by the projection effect, when the oscillation polarisation plane is almost parallel to the line of sight (LoS). In NOAA 11647 we could not identify any distinct loop structures visible at 171 \AA. 

The detected oscillations do not show any systematic decay. The duration of a single oscillation event is rather determined  by observational conditions  than by decay.
Consider several examples shown in Fig. \ref{fig_4}. In the loop shown in the first panel of the left column, the oscillation becomes unseen when the loop becomes too bright, perhaps because of another loop that moved in the LoS. In the third panel of the right column, the loop is seen to oscillate until its disappearance in this bandpass at time about 210 min. In the third panel of the left panel, again the oscillation is visible when the loop is seen in the panel, between 111 and 135 min. Thus, we  can speculate that the oscillations are generally undamped and are not detected when the host loop becomes invisible in the bandpass, or because of optically thin effects, when another bright coronal structure comes in the LoS, or, perhaps, because of the turn of the polarisation plane of the kink oscillation.

Lengths of the oscillating loops were estimated under the assumption of a three-dimensional semi-circular shape as $ L = \pi R$, where $R$ is the major radius of the loop measured either as the half distance between the loop's footpoints if both the footpoints were seen on the solar disk, or otherwise as the loop's height when the loop is seen off-limb.
Results of the analysis are shown in Table~\ref{table}. 
                         
\begin{figure}
\centering
\resizebox{\hsize}{!}{\includegraphics{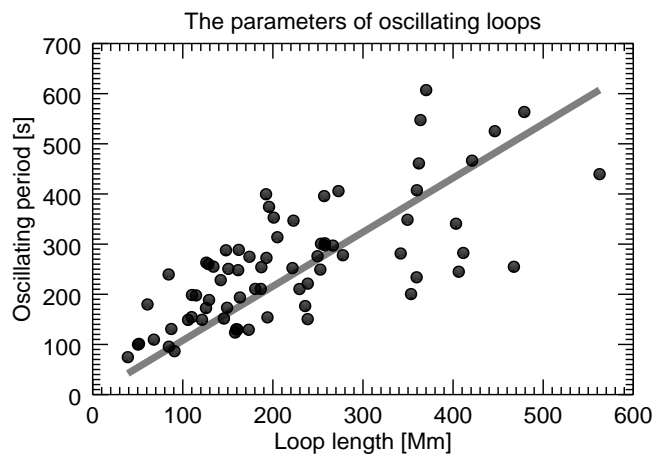}}
\caption{Dependence of the  oscillation periods on the loop lengths. Every circle corresponds to an oscillation event. The linear fit is shown by the solid line.}
\label{fig_5}
\end{figure}

The large number of the oscillating loops, precisely 72,  analysed in this study allows us to establish a statistically significant relationship between the loop length and the oscillation period. 
Figure~\ref{fig_5} shows that despite significant scattering of the data, the period increases with the length of the loop.
The Pearson correlation coefficient is as large as $r = 0.72 \pm 0.12$ with its standard deviation estimated as $1/\!\sqrt{N}$, where $N$ is the number of points.
In our case, $N = 72$, and thus $1/\!\sqrt{N}=0.12$, which means that the correlation coefficient has a 6-$\sigma$ significance.
The dependence was fitted with a linear function $P = (1.08 \pm 0.04)L$, which is shown with a solid line on Figure~\ref{fig_5}.
Here $P$ is the oscillation period measured in seconds and $L$ is the loop length measured in megameters.
                
\begin{figure*}
        \centering
        \resizebox{\hsize}{!}{\includegraphics{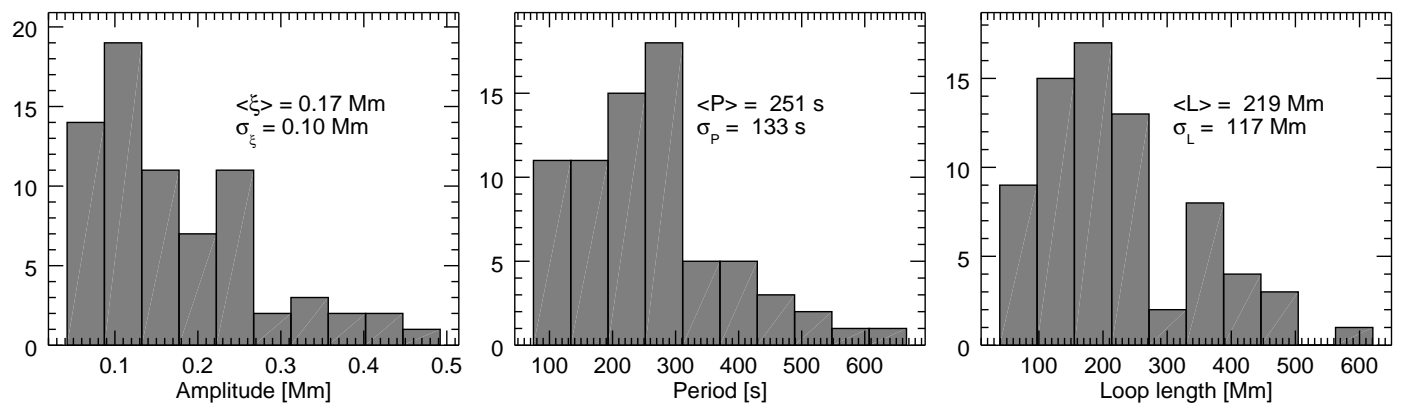}}
        \caption{\textbf{Distributions} of the parameters of the oscillating loops: amplitude (\textit{left panel}), period (\textit{middle panel}) and the length of the loop (\textit{right panel}).
         Average values ($<\xi>$, $<P>$ and $<L>$) and standard deviations ($\sigma_\xi$, $\sigma_P$ and $\sigma_L$) of the corresponding quantities are also provided.}
        \label{fig_6}
\end{figure*}

\begin{figure*}
        \centering
        \resizebox{\hsize}{!}{\includegraphics{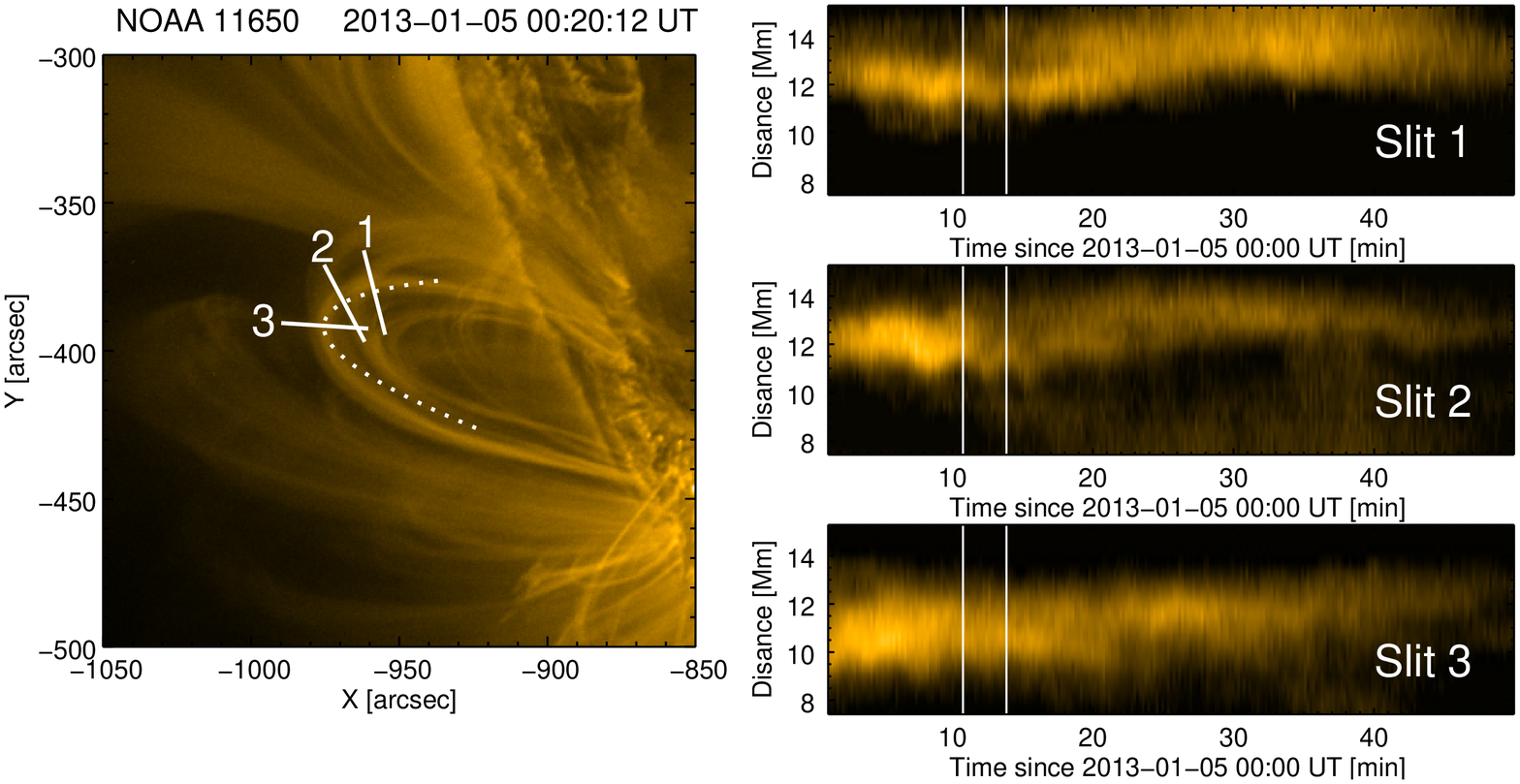}}
        \caption{Illustration of the in phase oscillation of different loop segments. The 171 \AA~ image of the active region NOAA 11650 taken on 5 January 2015 00:20:12 UT is presented in the \textit{left panel}.  The oscillated loop is highlighted with a dotted line. Bold white lines show three slits used for constructing time-distance plots. The time-distance plots for three  different positions are shown in the \textit{right panel}. One oscillation cycle is indicated with white vertical lines.  }
        \label{fig_7}
\end{figure*}

\section{Discussion and conclusion}
\label{con}

Our study reveals that low-amplitude decayless kink oscillations of coronal loops is a persistent feature of the solar corona. The oscillations were found for all loops with sufficiently contrast boundaries in all analysed active regions except two active regions with short or faint loops. Thus we conclude that this oscillatory regime is a common feature of coronal loops. In all the cases, the oscillations appear without any established relation with impulsive energy releases in the corona. Different segments of oscillating loops are displaced by the oscillation in phase.
A typical case is shown in Fig. \ref{fig_7}. The oscillations could only be resolved  in a relatively small loop segment  close to the top of the loop (between slits 1 and 3). The oscillation pattern is rather noisy. However we do not see any evidence of the propagating wave. 

The histograms of oscillation parameters such as amplitude, period, and the length of the oscillating loop, as well as their average values and standard deviations, are presented in Fig. \ref{fig_6}.
The average apparent displacement amplitude is 0.17~Mm, which agrees with previous results from \cite{2013A&A...560A.107A}, and with the recent estimates of transverse oscillations in polar plumes from \cite{2014ApJ...790L...2T}.  The oscillation periods are found to range from 1.5 to 10 minutes. No clear evidence of the vertical polarisation was found. Thus, parameters of the oscillations are consistent with our previous results for  decayless kink oscillations described in \citep{2013A&A...552A..57N,2013A&A...560A.107A}.

The periods of decayless kink oscillations have been observed in the range that includes 3 and 5 minutes \citep{2013A&A...560A.107A}, which could be associated with sunspot oscillations leaking into the corona \citep{2009A&A...505..791S}. The 
presence of sunspots in the active region has been verified by the visual inspection of SDO/HMI images, and is shown in the fourth column of Table~\ref{table}. There is not a clear association with decayless kink oscillations since these are found in the same proportion in both types of active regions, with or without a photospheric sunspot. This provides an indication that sunspots oscillations probably do not play a prominent role in the excitation of decayless kink oscillations. Moreover, our findings do not show any increase in the appearance of the oscillations with\ periods of three or five minutes. 

An important finding is the correlation of the oscillation period with the length of the oscillating loop. This result confirms the interpretation of  decayless kink oscillations as standing natural oscillations of coronal loops. Indeed, the period of a standing kink mode is about double the loop length divided by the kink speed. The scattering of the data in Fig.~\ref{fig_5} can be readily attributed to the scattering in the values of the kink speeds determined by the densities of the plasma and the magnetic field strengths inside and outside the loops and the uncertainties of
the loop length estimations.

The physical mechanism responsible for the appearance of  decayless low-amplitude oscillations of coronal loops remains unknown. However, the persistency of this regime established by our study suggests that the oscillations are continuously excited by some perpetual driver. The driver could be random or quasi-harmonic, but in the latter case it should be out of the resonance with the frequency of the kink mode. Together with the continuous excitation, the oscillations should be subject to continuous damping, e.g. by resonant absorption. Essentially, this reasoning coincides with the empirical model given by Eq.~(5) of \mbox{\citep{2013A&A...552A..57N}},
\begin{equation}
\frac{d^2 \xi}{d\,t^2} + \delta \frac{d\,\xi}{d\,t}+\Omega^2_\mathrm{K}\xi = f(t),
\end{equation}
where $\xi$ is the loop displacement at a certain hight, $\delta$ is the damping coefficient, $\Omega_\mathrm{K}$ is the natural frequency of the kink oscillation,
and $f(t)$ is the non-resonant external driving force produced, e.g. by the perpetual motion of the loop footpoints because of granulation. However, this excitation mechanism would produce horizontally and vertically polarised oscillations with the equal efficiency, which is not supported by our results. Thus, we conclude that the mechanism for the feeding the oscillations with energy remains unidentified.

\begin{acknowledgements}
The work was supported by the Marie Curie PIRSES GA-2011-295272 \textit{RadioSun} Project; the STFC consolidated grant ST/L000733/1 (GN, VMN), the European Research Council under the \textit{SeismoSun} Research Project No. 321141; the Russian Foundation of Basic Research under grants 13-02-00044-a, 15-02-01089-a, 15-02-03835-a, and 15-32-20504 mol\_a\_ved; the Federal Agency for Scientific Organisations base project II.16.3.2 "Non-stationary and wave processes in the solar atmosphere"; and the BK21 plus program through the National Research Foundation funded by the Ministry of Education of Korea (VMN).
\end{acknowledgements}


%
\begin{longtab}
        \centering                           
        \begin{longtable}{ c c c c c c c c } 
        \caption{\textbf{List of the analysed oscillating loops and parameters of the oscillations.}   
                \label{table}
                }
                 \\       
        \hline \hline                 
        \textbf{AR}& \textbf{Date}& \textbf{Oscillations}&  \textbf{Sunspots}& \textbf{Loop No}        & \textbf{Loop length}  & \textbf{Period }      & \textbf{Amplitude} \\
                        &               & yes/no        & yes/no        &       & [Mm]                    & [s]           & [Mm]          \\
        \hline
        \endfirsthead
        \caption{\textbf{List of the analysed oscillating loops and parameters of the oscillations (continued)}.}\\
        \hline \hline                 
                \textbf{AR}& \textbf{Date}& \textbf{Oscillations}&  \textbf{Sunspots}& \textbf{Loop No}        & \textbf{Loop length}  & \textbf{Period }      & \textbf{Amplitude} \\
                                &               & yes/no        & yes/no         &       & [Mm]                  & [s]           & [Mm]          \\
                \hline
                \endhead
                \hline
                \endfoot

        \multirow{2}{*}{11637}& \multirow{2}{*}{2013-01-05}     & \multirow{2}{*}{yes}  &\multirow{2}{*}{yes}   & 1                       &       190             &       250             & 0.09            \\
                        &       &       &       & 2*                    &       120                     &       160             & 0.18            \\ 
                \hline  
        11638   & 2013-01-07  & \textbf{no}&    yes& ...                &       ...                     &       ...     & ...             \\      
        \hline  
\multirow{5}{*}{11639}& \multirow{5}{*}{2013-01-04}     & \multirow{5}{*}{yes}  &\multirow{5}{*}{no}& 1                               &       150                     &       170             & 0.07            \\
                        &       &       &       & 2                     &       230                     &       210             & 0.13            \\
                        &       &       &       & 3*            &       130                     &       160             & 0.10            \\
                        &       &       &       &       4*              &       80                      &       270             & 0.14            \\
                        &       &       &       &       5               &       130                     &       190             & 0.05            \\      
                \hline          
\multirow{7}{*}{11640}& \multirow{7}{*}{2013-01-05}     & \multirow{7}{*}{yes}  &\multirow{7}{*}{yes}&  1                       &       200                     &       310             & 0.08            \\
                        &       &       &       &       2               &       190                     &       150             & 0.09            \\
                        &       &       &       &       3               &       180                     &       210             &         0.24    \\
                        &       &       &       &       4               &       250                     &       250             &       0.13    \\
                        &       &       &       &       5*              &       250                     &       250             &       0.30            \\
                        &       &       &       &       6               &       110                     &       200             &       0.24    \\ 
                        &       &       &       &       7*              &       200                     &       360             &       0.08    \\ 
                \hline
\multirow{3}{*}{11641}& \multirow{3}{*}{2013-01-11}     & \multirow{3}{*}{yes}  &\multirow{3}{*}{no}&   1                       &       110                     &       200             &       0.17    \\
                        &       &       &       &       2               &       200                     &       350             &       0.26    \\
                        &       &       &       &       3               &       370                     &       610             &       0.37    \\ 
                \hline          
        \multirow{8}{*}{11642}& \multirow{8}{*}{2012-12-31}     & \multirow{8}{*}{yes}  &\multirow{8}{*}{yes}&  1               &       420                     &       470             &         0.13    \\
                        &       &       &       &       1               &       420                     &       240             &       0.07    \\
                        &       &       &       &       2               &       250                     &       300             &       0.31    \\
                        &       &       &       &       3               &       160                     &       120             &       0.05    \\
                        &       &       &       &       4               &       280                     &       280             &       0.25    \\
                        &       &       &       &       5               &       350                     &       350             &       0.23    \\
                        &       &       &       &       6*              &       260                     &       300             &       0.33    \\
                        &       &       &       &       7               &       450                     &       530             &       0.33    \\ 
        \hline          
        \multirow{6}{*}{11643}& \multirow{6}{*}{2012-12-31}     & \multirow{6}{*}{yes}  &\multirow{6}{*}{yes}&  1               &       360                     &       460             &       0.42    \\
                        &       &       &       &       2               &       150                     &       250             &       0.15    \\
                        &       &       &       &       3               &       160                     &       290             &       0.26    \\
                        &       &       &       &       4               &       190                     &       270             &       0.19    \\
                        &       &       &       &       5               &       80                      &       100             &       0.06    \\
                        &       &       &       &       6*              &       360                     &       550             &       0.35    \\ 
        \hline          
        \multirow{4}{*}{11644, 11646}& \multirow{4}{*}{2013-01-12}      & \multirow{4}{*}{yes}    &\multirow{4}{*}{yes}&  1               &       160                     &       250             &       0.17    \\
                        &       &       &       &       2               &       110                     &       150             &       0.16    \\
                        &       &       &       &       3               &       90                      &       130             &       0.05    \\
                        &       &       &       &       4               &       110                     &       150             &       0.09    \\
        \hline          
\multirow{5}{*}{11645}& \multirow{5}{*}{2013-01-10}     & \multirow{5}{*}{yes}  &\multirow{5}{*}{yes}&  1                       &       340                     &       280             &       0.45    \\
                        &       &       &       &       2               &       260                     &       300             &       0.23    \\
                        &       &       &       &       3               &       220                     &       350             &       0.19         \\
                        &       &       &       &       4               &       160                     &       130             &       0.11    \\
                        &       &       &       &       5               &         160                     &       130             &       0.12    \\
        \hline
        11647   &       2013-01-04      & no    & no    &       ...             &       ...                     &       ...             &       ...     \\              
        \hline          
        \multirow{5}{*}{11648}& \multirow{5}{*}{2013-01-15}     & \multirow{5}{*}{yes}  &\multirow{5}{*}{no}&   1               &       220                     &       250             &       0.37    \\
                        &       &       &       &       2               &       190                     &       210             &       0.13    \\
                        &       &       &       &       3               &       160                     &       190             &       0.11    \\
                        &       &       &       &       4               &       150                     &       150             &       0.12    \\
                        &       &       &       &       5               &       270                     &       410             &       0.13    \\ 
        \hline
        \multirow{2}{*}{11649}& \multirow{2}{*}{2013-01-13}     & \multirow{2}{*}{yes}  &\multirow{2}{*}{yes}&  1               &       50                      &       100             &       0.12    \\
                        &       &       &       &       2               &       70                      &       110             &       0.04    \\
        \hline
        \multirow{5}{*}{11650}& \multirow{5}{*}{2013-01-05}     & \multirow{5}{*}{yes}  &\multirow{5}{*}{yes}&  1               &       560             &       440             &       0.43    \\
                        &       &       &       &       2               &       410                     &       280             &       0.12    \\
                        &       &       &       &       3               &       240                     &       220             &       0.20            \\
                        &       &       &       &       4               &       170                     &       270             &       0.25    \\
                        &       &       &       &       5               &       170                     &       130             &       0.11    \\ 
        \hline          
        \multirow{2}{*}{11651}& \multirow{2}{*}{2013-01-07}     & \multirow{2}{*}{yes}  &\multirow{2}{*}{no}&   1               &       130                     &       260             &       0.06    \\
                        &       &       &       &       2               &       130                     &       260             &       0.16    \\      
        \hline          
        \multirow{6}{*}{11652}& \multirow{6}{*}{2013-01-07}     & \multirow{6}{*}{yes}  &\multirow{6}{*}{yes}&  1               &       400                     &       340             &       0.20            \\
                        &       &       &       &       2               &       240                     &       180             &       0.12    \\
                        &       &       &       &       3               &       470                     &       250             &       0.10            \\
                        &       &       &       &       4               &       360                     &       410             &       0.25    \\
                        &       &       &       &       5               &       240                     &       150             &       0.05    \\
                        &       &       &       &       6               &       260                     &       400             &       0.14    \\
        \hline          
        \multirow{2}{*}{11653}& \multirow{2}{*}{2013-01-06}     & \multirow{2}{*}{yes}  &\multirow{2}{*}{yes}&  1               &       90                      &       90              &       0.05    \\
                        &       &       &       &       2               &       140                     &       230             &       0.09    \\
        \hline          
        \multirow{4}{*}{11654}& \multirow{4}{*}{2013-01-08}     & \multirow{4}{*}{yes}  &\multirow{4}{*}{yes}&  1               &       360                     &       230             &       0.19    \\
                        &       &       &       &       2               &       350                     &       200             &       0.11    \\
                        &       &       &       &       3               &       480                     &       560             &       0.35    \\
                        &       &       &       &       4               &       270                     &       300             &       0.21    \\
        \hline          
        11655& 2013-01-14       & yes   &       yes&    1               &       50                      &       100             &       0.04    \\
        \hline
        \multirow{2}{*}{11656}& \multirow{2}{*}{2013-01-20}     & \multirow{2}{*}{yes}  &\multirow{2}{*}{no}&   1               &       60                      &       180             &       0.06    \\
                        &       &       &       &       2               &       40                      &       70              &       0.06    \\ 
        \hline          
        \multirow{3}{*}{11657}& \multirow{3}{*}{2013-01-20}     & \multirow{3}{*}{yes}  &\multirow{3}{*}{no}&   1               &       150                     &       290             &       0.25    \\
                        &       &       &       &       2               &       130                     &       260             &       0.15    \\
                        &       &       &       &       3               &       190                     &       400             &       0.23    \\
        \end{longtable}
        \tablefoot{The NOAA numbers of the active regions are listed in the first column. The date of the observation is given in second column. For each active region we show the presence or absence of kink oscillations (third column) and sunspots (forth column). Estimated lengths of the oscillating loops are given in the sixth column. The period and amplitude of the identified oscillation are given in the last two columns. The loops, oscillations of which are given in Fig.~\ref{fig_4} are marked by the asterisks.}
\end{longtab}

\bibliographystyle{aa} 
\bibliography{references} 

\end{document}